\documentclass[%
 reprint,
superscriptaddress,
 amsmath,amssymb,
 aps,
 prl,
]{revtex4-1}

\usepackage{graphicx}
\usepackage{dcolumn}
\usepackage{bm}
\usepackage{hyperref}
\hypersetup{
	colorlinks=true,
	linkcolor=blue,
	filecolor=blue,      
	urlcolor=blue,
	citecolor=blue,
}

\begin{document}

\preprint{APS/123-QED}
\title{Correlated Energy Spread Compensation in Multi-Stage Plasma-Based Accelerators}%

\author{A. Ferran Pousa}
 \email{angel.ferran.pousa@desy.de}
 \affiliation{Deutsches Elektronen-Synchrotron DESY, 22607 
 	Hamburg, Germany }
 \affiliation{Institut f\"ur Experimentalphysik, Universit\"at Hamburg, 22761 
 	Hamburg, Germany  }%
\author{A. Martinez de la Ossa}
\affiliation{Deutsches Elektronen-Synchrotron DESY, 22607 
	Hamburg, Germany }
\author{R. Brinkmann}
\affiliation{Deutsches Elektronen-Synchrotron DESY, 22607 
	Hamburg, Germany }
\author{R. W. Assmann}
\affiliation{Deutsches Elektronen-Synchrotron DESY, 22607 
	Hamburg, Germany }

\date{\today}

\begin{abstract}
The extreme electromagnetic fields sustained by plasma-based accelerators allow for energy gain rates above 100 GeV/m but are also an inherent source of correlated energy spread. This severely limits the usability of these devices. Here we propose a novel compact concept which compensates the induced energy correlation by combining plasma accelerating stages with a magnetic chicane. Particle-in-cell and tracking simulations of a particular 1.5 m-long setup with two plasma stages show that 5.5 GeV bunches with a final relative energy spread of $1.2\times10^{-3}$ (total) and $5.5\times10^{-4}$ (slice) could be achieved while preserving sub-micron emittance. This at least one order of magnitude below current state-of-the-art and paves the way towards applications such as Free-Electron Lasers.
\end{abstract}

\maketitle



Plasma-based accelerators (PBAs), driven either by charged particle beams (plasma wakefield accelerator, PWFA \cite{PhysRevLett.54.693}) or intense laser pulses (laser wakefield accelerator, LWFA \cite{tajima1979laser}), are able to sustain accelerating gradients in excess of 100 GeV/m \cite{faure2006controlled}. These extreme gradients are orders of magnitude higher than those achievable with radiofrequency technology and offer a path towards miniaturized particle accelerators with ground-breaking applications in science, industry and medicine \cite{malka2008principles}.

Steady progress over the past decades has led to the successful demonstration of electron bunches with multi-GeV energy \cite{leemans2006gev, wang2013quasi, leemans2014multi,mirzaie2015demonstration}, micron-level emittance \cite{fritzler2004emittance,brunetti2010low} and kiloampere current \cite{lundh2011few,couperus2017demonstration}. However, the high amplitude and short wavelength ($\sim 100 \ \mathrm{\mu m}$) of the wakefields naturally imprint a longitudinal energy correlation (or chirp) along the accelerated (witness) bunch, leading to a large relative energy spread typically on the 1-10 \% range \cite{ferranpousa2018limitations}. This is a long-standing issue for PBAs which critically impacts the beam quality \cite{migliorati2013}, particularly for applications such as Free-Electron Lasers (FELs) \cite{madey1971stimulated} where a relative energy spread $\lesssim 0.1\%$ is required \cite{corde2013femtosecond}.

Solving this issue is therefore key for demonstrating the usability of these devices. A well known concept for mitigating the correlated energy spread is that of beam loading \cite{katsouleas1987beam,doi:10.1063/1.1889444, tzoufras2008beam}, in which the witness bunch itself is used to flatten the slope of the accelerating fields. This, however, relies on a very precise shaping of the current profile and has yet to be demonstrated with the desired performance. Furthermore, since the optimal profile depends on the wakefield structure, a certain energy spread will always develop in LWFAs, where the wakefield experienced by the bunch will change due to the laser evolution \cite{esarey2009physics} as well as dephasing \cite{schroeder2011nonlinear}. Alternative ideas have also been proposed in order to achieve, in average, a flat accelerating gradient. These include modulating \cite{brinkmann2017chirp} or tailoring \cite{PhysRevLett.121.074802} the plasma density profile as well as injecting a secondary bunch \cite{manahan2017single}, but they show limited success or remain to be experimentally realized. A different approach contemplates stretching the bunch in order to minimize the slice energy spread \cite{maier2012demonstration}.

\begin{figure*}
	\includegraphics[width=\textwidth]{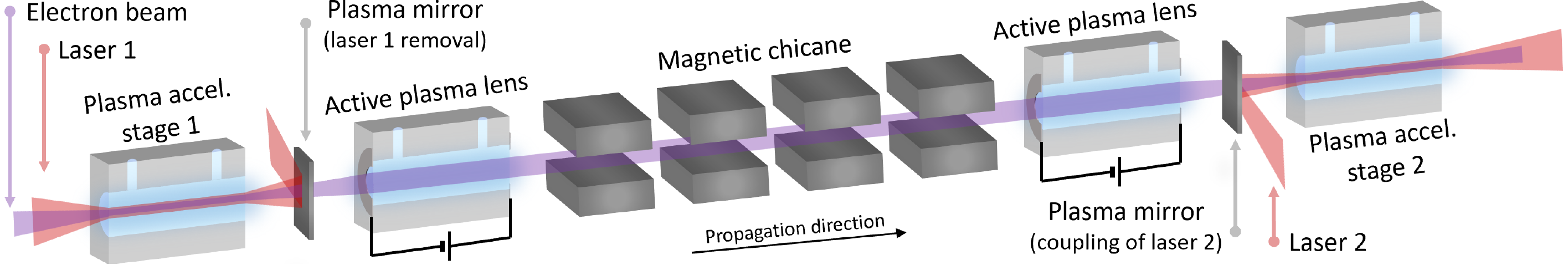}
	\caption{\label{fig:overview}Overview of the proposed accelerator concept (not to scale).}
\end{figure*}

In this Letter we propose a novel concept for compensating the correlated energy spread by taking advantage of the naturally occurring energy chirp. The scheme, illustrated in Fig. \ref{fig:overview}, consists mainly on two identical plasma accelerating stages joined by an intermediate magnetic chicane in which the longitudinal energy correlation of the bunch is inverted. Thus, the energy chirp generated in the first stage is compensated in the second. Numerical simulations with the Particle-in-Cell (PIC) code FBPIC \cite{LEHE201666} as well as the tracking codes ASTRA \cite{floettmann2011astra} and CSRtrack \cite{DOHLUS2000338} show that multi-GeV beams with unprecedented energy spread could be obtained with this method. Although LWFA stages are used here, the core idea behind the scheme would be equally valid in the case of a particle driver.

In order to introduce this concept, the blowout regime \cite{lotov2004blowout, lu2006nonlinear} of plasma acceleration will be considered. In this case the laser or beam driver is able to expel all background plasma electrons, leaving behind an ion cavity with uniform focusing gradient, $K=(m/2ec)\omega_p^2$, and an approximately constant longitudinal electric field slope, $E_z' \equiv \partial_zE_z\simeq (m/2e)\omega_p^2$, along most of the accelerating phase. Here $\omega_p = \sqrt{n_pe^2/m \epsilon_0}$ is the plasma frequency, $e$ and $m$ the electron charge and mass, $\epsilon_0$ the vacuum permittivity and $n_p$ the unperturbed plasma density. In order to describe the position and energy of the particles along the accelerator it is also useful to introduce the speed-of-light coordinate, $\xi = z - ct$, as well as the relativistic Lorentz factor, $\gamma = 1/\sqrt{1-(\bm{v}/c)^2}$, where $t$ is the time and $\bm{v}$ and $z$ are, respectively, the particle velocity and longitudinal position in the laboratory frame. Within the generated cavity, electrons perform transverse oscillations (known as betatron motion) with a frequency $\omega_\beta(t)=\sqrt{eK/m\gamma(t)}$, while their energy evolves as $\gamma(t) = \gamma_0 - (e/mc)E_zt$. 

For a particle bunch with average energy $\bar{\gamma}(t)=\left<\gamma(t)\right>$ centered at $\bar{\xi}$, the longitudinal chirp can be expressed as $\chi(t)=\left<\Delta\xi\Delta\gamma(t)\right>/\left<\Delta\xi^2\right>\bar{\gamma}(t)$, where $\Delta\gamma(t)=\gamma(t)-\bar{\gamma}(t)$ and $\Delta\xi=\xi-\bar{\xi}$. A simple expression for the chirp evolution within a plasma stage can be obtained if a constant $E_z'$ is assumed. This yields 
\begin{equation}
\chi(t)=\left(\chi_0\bar{\gamma}_0-\frac{e}{mc}E_z't\right)\bar{\gamma}(t)^{-1} \ ,
\end{equation}
which tends asymptomatically to $\chi=E_{z}'/E_{z}$ and where $\chi_0$ and $\bar{\gamma}_0$ are the initial bunch chirp and energy. If the bunch length is $\sigma_z=\sqrt{\langle \Delta\xi^2 \rangle}$, this induces a correlated energy spread $\sigma^\mathrm{corr}_\gamma(t)/\bar{\gamma}(t)=\chi(t)\sigma_{z}$. In a 2-stage accelerator as in Fig. \ref{fig:overview}, the accumulated chirp after a first stage of length $L_{p,1}$ for an initially unchirped bunch will be $\chi_1=-(e/mc^2)E_{z,1}'L_{p,1}/\bar{\gamma}_1$. Therefore, if the longitudinal phase space of the bunch is inverted at this point such that $\hat{\chi}_1=-(\sigma_{z,1}/\sigma_{z,2})\chi_1$ is obtained, the correlated energy spread could be compensated in a following stage fulfilling $E_{z,2}'L_{p,2}=(mc^2/e)\hat{\chi}_1\bar{\gamma}_1$. For a symmetric inversion ($\sigma_{z,1}=\sigma_{z,2}$), using two identical plasma stages (same $E_z'$ and $L_p$) would be the simplest setup.

\begin{figure}[!htb]
	\centering
	\includegraphics*[width=240pt]{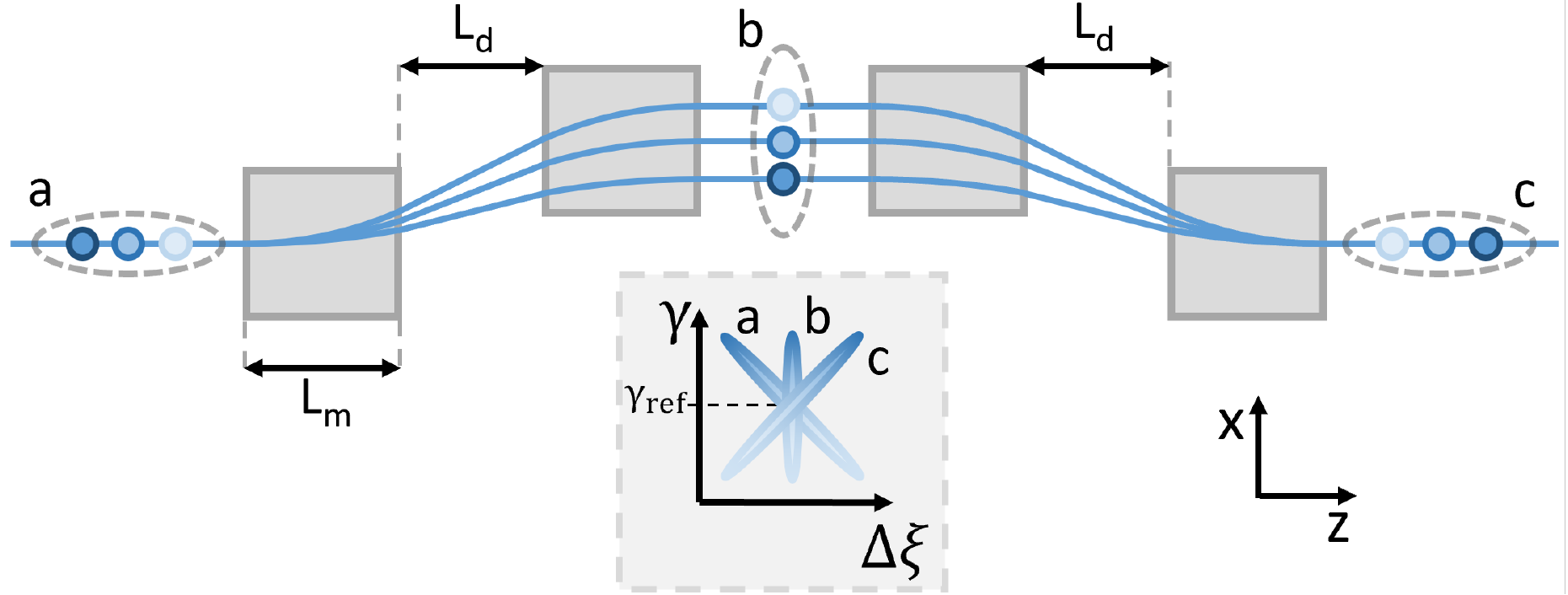}
	\caption{Working principle of a magnetic chicane with $R_{56}=2/\chi$. The bunch longitudinal phase space is shown at the chicane entrance (a), middle (b) and exit (c). Darker color implies higher energy.} 
	\label{fig:chicane_sketch}
\end{figure}

This longitudinal phase space inversion can be performed with a conventional chicane. As illustrated in Fig. \ref{fig:chicane_sketch}, this device is composed by 4 dipole magnets in which particles undergo an energy-dependent trajectory bend. With respect to a hypothetical reference particle with $\gamma=\gamma_\mathrm{ref}$, those with $\gamma>\gamma_\mathrm{ref}$ experience less bending and therefore a shorter path length, while the opposite occurs for those with $\gamma<\gamma_\mathrm{ref}$. Defining $\delta=(\gamma-\gamma_\mathrm{ref})/\gamma_\mathrm{ref}$, the path length differences after the chicane, $\Delta\xi_\mathrm{ch}$, can be expressed with respect to the reference particle as
\begin{equation}\label{eq:path_length_chic}
\Delta\xi_\mathrm{ch}(\delta)= R_{56} \delta + T_{566} \delta^2 + \mathcal{O}(\delta^3)  \ ,
\end{equation}
where $T_{566}\simeq-3/2R_{56}$ \cite{chao1999handbook}. To first order, the $R_{56}$ coefficient can be simply determined as $R_{56}=\Delta\xi_\mathrm{ch} / \delta=\Delta\xi_\mathrm{ch} / \chi \Delta\xi$ (assuming $\gamma_\mathrm{ref}=\bar{\gamma}$). These path length differences allow for a certain control of the longitudinal phase space. For example, full bunch compression can be achieved if $\Delta\xi_\mathrm{ch}$ exactly compensates the initial offsets with respect to the bunch center, $\Delta\xi$. Similarly, inverting the chirp, i.e inverting the bunch along $\xi$, can be achieved if $\Delta\xi_\mathrm{ch} = 2 \Delta\xi$. This implies that a chicane with $R_{56} = 2/\chi$ is required.

However, successfully performing this process requires $\chi$, and therefore $E_z'$, to be uniform along the bunch. Thus, this scheme is ideally suited for weakly beam-loaded wakefields, where $E_z'$ is not perturbed by the bunch, or where the beam-loading effect linearly modifies the accelerating fields \cite{tzoufras2008beam}. Additionally, in order to maintain a purely linear chirp after the chicane, the higher order terms in Eq. (\ref{eq:path_length_chic}) need to be minimized. This is required because the non-linear contributions cannot be compensated in the second stage and therefore will lead to increased energy spread.  In particular, keeping the higher order effects below 1\%, i.e $|T_{566}\delta/R_{56}| \lesssim 10^{-2} $, requires $\sigma_z \lesssim  0.015 c/\omega_p$ if typical blowout fields are assumed ($E_z' = (m/2e)\,\omega_p^2$ and $E_z = (mc/e)\omega_p$). These contributions can also be mitigated by sextupole magnets \cite{PhysRevSTAB.8.012801} or by imprinting non-linearities in the bunch to compensate those from the chicane, e.g., by optimizing a non-uniform $E_z'$ along the bunch. Still, for a non-uniform $\chi$, the scheme could be designed to mitigate the chirp at the bunch core in order to minimize the slice energy spread for applications such as FELs.

Once the chicane $R_{56}$ is determined, the magnet length $L_m$ and bending angle $\theta$ experienced by the reference particle can be directly determined from $R_{56} = -2\theta^2 (L_d + 2L_m/3)$ \cite{chao1999handbook},
where $L_d$, as defined in Fig. \ref{fig:chicane_sketch}, is the length of the drift space between the first and second as well as the third and fourth dipoles. The magnetic field strength can then be obtained as $B=(mc/e)\theta\gamma_\mathrm{ref}/L_m$. Assuming a bunch with $\chi=E_{z}'/E_z$ and typical blowout fields, it can be obtained that $R_{56}\simeq4c/\omega_p\ll 1 \ \mathrm{mm}$ (for $n_p\gtrsim10^{16} \ \mathrm{cm^{-3}}$) which, considering $L_m\sim L_d\sim0.1 \ \mathrm{m}$ leads to $\theta \ll 0.1 \ \mathrm{rad}$. The high $\chi$ characteristic of PBAs therefore allows for a very compact chicane design ($\sim 1$ m) while requiring a very small bending angle. This greatly minimizes the impact of Coherent Synchrotron Radiation (CSR) \cite{DOHLUS1997494} on the beam parameters.

A possible implementation of this scheme is shown in Fig. \ref{fig:overview}. Two LWFA stages accelerating an externally injected bunch are joined by a magnetic chicane, including active plasma lenses (APLs) \cite{PhysRevLett.115.184802} to transport the beam and plasma mirrors \cite{thaury2007plasma} to couple the laser pulses in and out. The two required laser drivers could originate from splitting a single original pulse and therefore be intrinsically synchronized. 

APLs consist on circular, gas-filled capillaries with typically sub-mm radius, $R_c$, on which a multi-kV discharge is applied by electrodes at both ends, causing a breakdown of the gas. A current is then driven though the ionized plasma, generating radially symmetric focusing fields with kT/m gradients which are linear up to a radius $r\lesssim R_c/2$ \cite{PhysRevLett.115.184802}. The strong focusing fields make these devices ideal for transporting the strongly divergent and high energy spread bunches coming out of the first stage, as they can be placed very close to it, thus mitigating emittance growth in the drift \cite{migliorati2013}, and they can focus on both planes, which significantly reduces their chromaticity with respect to other focusing systems \cite{PhysRevLett.115.184802}. 

The plasma mirrors are necessary in order to maintain the compact footprint of the design, as they can be placed close to the laser focal spot. In comparison, conventional mirrors would have to be placed several meters away in order to avoid damage from the multi-TW lasers required by LWFAs. As an example, tape-based plasma mirrors have been successfully used to remove and couple in laser pulses \cite{steinke2016multistage}. However, due to their thickness, they could also perturb the witness bunch due to scattering as well as the plasma created by the incident laser. A promising alternative are liquid crystal films \cite{poole2016experiment}, which can have a thickness down to the few-nm scale. A specific implementation is not considered here.

This scheme, apart from offering an energy spread compensation, could also reduce the sensitivity to critical issues such as the timing jitter between laser driver and witness bunch \cite{1742-6596-874-1-012032} or the hosing instability  \cite{PhysRevLett.67.991}. The timing jitter is one of the main challenges of external injection, as it translates into a large energy jitter at the LWFA exit due to the large $E_z'$. However, thanks to the chicane in this scheme, an injection offset with respect to the ideal phase in the first LWFA would translate, to first order in Eq. (\ref{eq:path_length_chic}), into the opposite offset at the second LWFA, thus providing a stable average accelerating field and energy output. Furthermore, since the bunches are accelerated with a large energy chirp, the hosing instability is also mitigated \cite{PhysRevLett.119.244801, PhysRevLett.118.174801}.

In order to test the performance of this scheme, start-to-end simulations for a particular set of parameters have been performed. The LWFA stages and APLs have been simulated using the spectral, quasi-3D PIC code FBPIC, while the tracking code ASTRA has been used for the remaining beamline elements taking into account 3D space-charge effects. Additionally, CSRtrack has also been used to account for CSR effects in the chicane.

Motivated by the parameters from the EuPRAXIA design study \cite{1742-6596-874-1-012029}, the simulated setup aims at providing 5 GeV electron beams suitable for FEL applications, i.e., a peak current in the kA range, sub-micron emittance and, specially, an energy spread $\lesssim 0.1\%$. For this purpose, we consider an externally injected Gaussian electron bunch with an initial energy of 250 MeV, a 0.5\% energy spread with no chirp, a normalized transverse emittance $\epsilon_{n,x} = 0.5 \ \mathrm{\mu m \, rad}$, 10 pC of charge, a FWHM duration $\tau=5 \ \mathrm{fs}$ and a peak current $I_\mathrm{peak}\simeq 2 \ \mathrm{kA}$. The bunch transverse size, $\sigma_x$, is matched \cite{assmann1998transverse, PhysRevSTAB.15.111303} to the plasma focusing fields in order to prevent emittance growth. This requires the beam beta function, $\beta_x = \gamma\sigma_x^2/\epsilon_{n,x}$, to satisfy $\beta_x=c/\omega_\beta$ at the LWFA entrance. The normalized emittance is defined as $\epsilon_{n,x}=(\langle x^2\rangle \langle p_x^2\rangle -\langle x p_x \rangle^2)^{1/2}/mc$, where $p_x$ is the transverse particle momentum. Electron bunches within this range of parameters can be produced with conventional accelerators \cite{zhu2018simulation}. The two identical LWFAs have a length $L_p=8 \ \mathrm{cm}$ and a parabolic transverse density profile $n_p = n_{p,0} + r^2/\pi r_e w_0^4$, where $n_{p,0}=10^{17} \ \mathrm{cm}^{-3}$ is the on-axis plasma density, $r$ the radial coordinate, $r_e$ the classical electron radius and $w_0$ the spot size of the laser driver. Plasma cells in this range of parameters have been recently demonstrated \cite{PhysRevE.97.053203} and provide ideal guiding properties for the driving laser pulse. For simplicity, a longitudinal flat-top plasma density profile has been considered, although the presence of smooth plasma-to-vacuum transitions would be beneficial for electron beam matching \cite{PhysRevSTAB.17.054402} and emittance growth minimization \cite{PhysRevSTAB.18.041302}. This choice of electron beam and plasma parameters also helps in reducing below the $10^{-4}$ level the energy spread generated in the LWFAs due to slice mixing from betatron motion \cite{pousa2018limitations}. Each LWFA is driven by a 40 J, 0.75 PW laser pulse with a peak normalized vector potential $a_0=3$, a spot size $w_0=50 \ \mathrm{\mu m}$ and a FWHM duration $\tau_{0}=50 \ \mathrm{fs}$. This laser can be successfully guided throughout the LWFAs, which provide an energy gain of $\sim 2.6$ GeV each. The APLs are placed 3 cm away from the LWFAs, leaving enough space for the plasma mirrors. They provide a focusing gradient of 3 kT/m, as measured experimentally \cite{PhysRevLett.115.184802}, and have a length of 6.6 cm, optimized to achieve a beam waist at the chicane center. A plasma density $n_p^\mathrm{APL}=10^{15} \ \mathrm{cm^{-3}}$ has been considered in order to minimize the impact of beam-driven wakefields \cite{lindstrom2018analytic}. The chicane has a total length of 1.2 m, with $L_d = 12.5 \ \mathrm{cm}$ and dipoles with $L_m = 20 \  \mathrm{cm}$ and $B=0.54 \ T$ for a bending angle of $\theta=0.011 \ \mathrm{rad}$. 

\begin{figure}[!htb]
	\centering
	\includegraphics*[width=240pt]{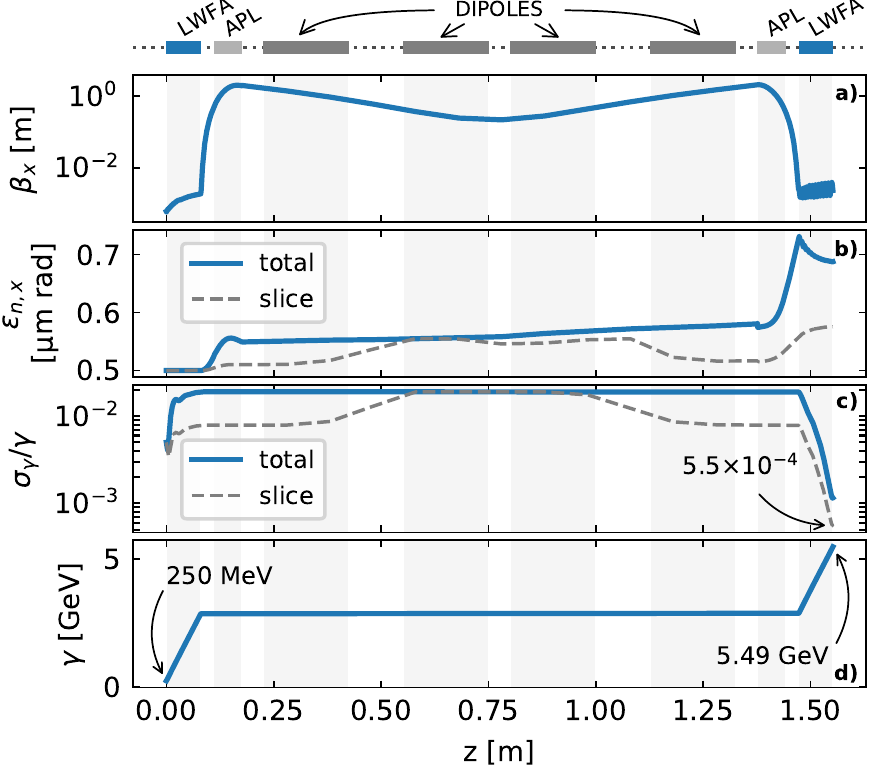}
	\caption{Evolution of the beta function (a), normalized emittance (b), energy spread (c) and energy (d) along the beamline. The emittance and beta function in the chicane are corrected for dispersion up to second order.} 
	\label{fig:10pC_results}
\end{figure}

An overview of the simulation results can be seen in Fig. \ref{fig:10pC_results}. The electron bunch leaves the first LWFA with preserved emittance, an energy of $\sim2.9$ GeV and a chirp $\chi\simeq-0.031 \ \mathrm{\mu m}^{-1}$, which induces a total relative energy spread $\sim 2\%$. As a consequence, the projected emittance grows after the accelerating stage until the beam divergence is controlled by the APL. The maximum beam size within the lens is $\sigma_r\simeq15 \ \mathrm{\mu m}\ll R_c/2$ and therefore experiences linear focusing. ASTRA and CSRtrack simulations show that the influence of space-charge and CSR on the beam parameters is negligible thanks to its GeV energy and the small bending angle. The beam is then focused in the following APL and injected into the second LWFA, where it gains an additional $\sim 2.6 \ \mathrm{GeV}$ for a final energy of $\sim 5.5 \ \mathrm{GeV}$ while reducing its relative energy spread down to $0.12 \%$.

\begin{figure}[!htb]
	\centering
	\includegraphics*[width=240pt]{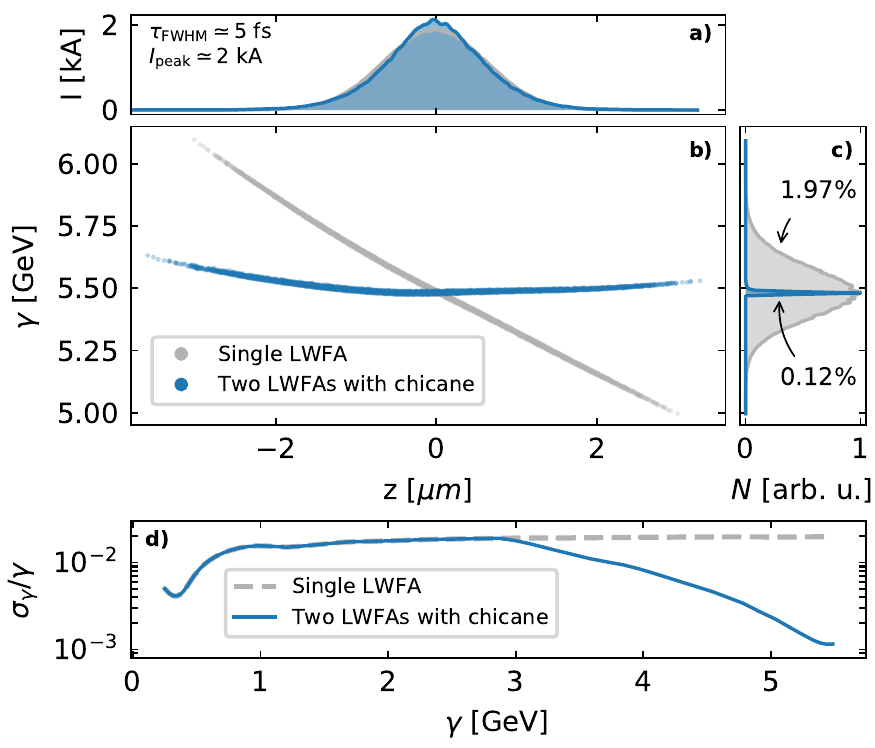}
	\caption{Comparison of the final bunch properties with respect to a case with acceleration in a single LWFA (same driver and plasma profile but $\sim 20$ cm long). The bunch current (a), longitudinal phase space (b) and energy profile (c) normalized to the peak number of counts, $N$, are shown. The total energy spread evolution during acceleration can be seen in (d).} 
	\label{fig:10pC_comparison}
\end{figure}

As seen in Fig. \ref{fig:10pC_comparison}, the relative energy spread is reduced by a factor $\sim20$ with respect to a case with no chirp compensation. However, it is only a factor $\sim5$ smaller than the value at injection, while the energy has increased more than 20 times. This is due the non-linear contributions in the chicane, as seen in Fig. \ref{fig:10pC_comparison}b, which increase the final projected energy spread. 

The emittance evolution in the drifts and APLs is well controlled, achieving a final value of $0.69  \ \mathrm{\mu m \, rad}$ ($\sim 37\%$ increase). This growth arises from the large energy chirp along the bunch, which causes individual slices to diverge (or converge) at different rates in the drifts and to have a different betatron frequency in the APLs, leading to increased projected emittance. However, it should be noted that this growth is not the same in both APLs, but it is more moderate in the first one. This is due to the beam-induced focusing wakefields which, for a short bunch, grow linearly towards the back \cite{lindstrom2018analytic} and, for a case with $\chi<0$, can mitigate the projected emittance growth by equalizing $\omega_\beta$ along the bunch. This suggests that wakefields in APLs, which are typically regarded as a key limitation of these devices \cite{lindstrom2018analytic}, can also be useful and could be optimized for emittance preservation in bunches with a large negative energy chirp. Another consequence of this slice decoherence is that $\beta_x$ will evolve differently along the bunch. Therefore, not all slices will be matched to the focusing fields in the second LWFA, causing the oscillations in $\beta_x$ seen in Fig. \ref{fig:10pC_results}a.

Regarding the slice parameters, which are key for FEL applications, this scheme achieves a final average relative energy spread and emittance of $5.5\times 10^{-4}$ and $0.58  \ \mathrm{\mu m \, rad}$, respectively, assuming $1 \ \mathrm{\mu m} $ slices. This energy spread is at least one order of magnitude lower than in state-of-art LWFAs and would satisfy the requirements for an X-ray FEL. The growth of the slice parameters in the chicane arises from the reduction in bunch length, which reaches its minimum at the chicane center and becomes shorter than a single slice. Thus, the slice parameters approach those of the whole bunch.

The presented scheme therefore offers a path towards ultra-low energy spread ($\sim10^{-4}$) beams while reducing the sensitivity to issues such as the timing jitter for external injection and the hosing instability. The concept is ideally suited for PBAs with weakly or linearly beam-loaded wakefields. Electron beams obtained with this method could enable ground breaking applications, such as compact FELs, and could be accelerated to higher energies thanks to its modular design, allowing other applications such as plasma-based colliders. This could be achieved either by introducing additional plasma stages while keeping a single central chicane, or by repeating multiple sections of one chicane every two plasma accelerating modules. Furthermore, this scheme could also be modified as an extreme bunch compressor by halving the chicane $R_{56}$. CSRtrack simulations with the same bunch produced in the first LWFA show that ultra-short bunches with a duration down to $\sim 70$  attoseconds and a $\sim 50 \ \mathrm{kA}$ peak current could be obtained, with possible applications in ultrafast science \cite{zewail2003atomic}.

\begin{acknowledgments} 
	This project has received funding from the European Union's 
	Horizon 2020 research and innovation programme under grant agreement No. 
	653782. We acknowledge the use of the High-Performance Cluster (Maxwell) at DESY for the numerical simulations. We would like to thank T. Mehrling, S.K. Barber, S. Steinke  and M. Dohlus for useful discussions.
\end{acknowledgments} 

\bibliography{multistage_compensation}

\end{document}